\documentclass[12pt]{iopart}
\usepackage{dsfont}
\usepackage{verbatim}
\usepackage{graphicx}
\usepackage{epsfig}
\usepackage{amssymb}
\usepackage{dcolumn}
\usepackage{bm}
\usepackage{braket}
\usepackage{mathtools}
\usepackage[nice]{nicefrac}
\usepackage{bbm}
\usepackage[usenames,dvipsnames]{xcolor}
\usepackage[colorlinks=true,citecolor=MidnightBlue,linkcolor=MidnightBlue,urlcolor=MidnightBlue]{hyperref}
\usepackage{siunitx}
\usepackage{cite}
\usepackage[normalem]{ulem}

\definecolor{Light Salmon Pink}{RGB}{255, 154, 162}
\definecolor{Melon}{RGB}{255, 183, 178}
\definecolor{Very Pale Orange}{RGB}{255, 218, 193}
\definecolor{Dirty White}{RGB}{226, 240, 203}
\definecolor{Magic Mint}{RGB}{181, 234, 215}
\definecolor{Crayola's Periwinkle}{RGB}{199, 206, 234}
\definecolor{Dark Pastel Red}{RGB}{194, 59, 35}
\definecolor{Carrot Orange}{RGB}{243, 154, 39}
\definecolor{Minion Yellow}{RGB}{234, 218, 82}
\definecolor{Dark Pastel Green}{RGB}{3, 192, 60}
\definecolor{Silver Lake Blue}{RGB}{87, 154, 190}
\definecolor{Dark Pastel Purple}{RGB}{151, 110, 215}

\definecolor{matblue}{RGB}{0, 113.9850, 188.9550}
\definecolor{matred}{RGB}{216.7500, 82.8750, 24.9900}
\definecolor{matyellow}{RGB}{236.8950, 176.9700, 31.8750}
\definecolor{matpurple}{RGB}{125.9700, 46.9200, 141.7800}
\definecolor{matgreen}{RGB}{118.8300, 171.8700, 47.9400}
\definecolor{matcyan}{RGB}{76.7550, 189.9750, 237.9150}

\PassOptionsToPackage{numbers,sort&compress}{natbib}


\colorlet{BLUE}{blue}

\newcommand{\mb}{\mathbf}

\begin{document}

\newcommand*{\QOQI}{Quantum Optics and Quantum Information Group, Friedrich-Alexander-Universität Erlangen-Nürnberg (FAU), Staudtstr. 1, 91058 Erlangen, Germany}

\newcommand*{\Health}{Health Science Core Facilities, University of Utah, Salt Lake City, UT 84112, United States of America}
\newcommand*{\Metrologie}{Institut für Metrologie, Freie Universität Berlin,Carl-Heinrich-Becker-Weg 6-10, 12165 Berlin, Germany}
\newcommand*{\Laser}{Department of Physics, Friedrich-Alexander-Universität Erlangen-Nürnberg (FAU), Staudtstr. 1, 91058 Erlangen, Germany}

\newcommand*{\SAOT}{Erlangen Graduate School in Advanced Optical Technologies (SAOT), Friedrich-Alexander Universit\"at Erlangen-N\"urnberg, Paul-Gordan-Str. 6, 91052 Erlangen, Germany}

\newcommand*{\qoqi}{Quantum Optics and Quantum Information Group, Friedrich-Alexander-Universität Erlangen-Nürnberg, Staudtstr. 1, 91058 Erlangen, Germany}

\title[Quantum Signatures of Two-Electron HBT Interference in Free Space]{Quantum Signatures of Two-Electron Hanbury Brown–Twiss Interference in Free Space}


\author{Florian~Fleischmann$^{1,*}$, Mona~Bukenberger$^{2}$, Anton Classen$^{3}$, Marc-Oliver Pleinert$^{1}$, and Joachim von Zanthier$^{1}$}

\address{$^1$ \QOQI}
\address{$^2$ \Metrologie}
\address{$^3$ \Health}
\vspace{10pt}
\begin{indented}
\item[] $^*$E-Mail: flo.fleischmann@fau.de
\end{indented}

\begin{abstract}
Understanding how fermionic exchange and Coulomb repulsion jointly shape two-electron correlations is essential for identifying genuine quantum signatures in multi-electron interference experiments.
To address this interplay, we investigate Hanbury Brown and Twiss interference of two electrons generated by two independent needle-tip emitters within a full quantum-mechanical framework.
In the absence of Coulomb interaction, the approach reproduces the results previously obtained within a quantum path formalism. 
For Coulomb-interacting electrons, we predict characteristic features absent in a semiclassical description: a pronounced Coulomb-dominated suppression region as well as Coulomb-induced phase offsets and fringe shifts. 
At the same time, outside of the Coulomb-dominated region, the spatial oscillation frequency is essentially governed by fermionic exchange symmetry.
Our results establish quantitative parameter regimes for disentangling Coulomb interaction from fermionic exchange symmetry in such experiments.

\end{abstract}

\section{Introduction}

The quantum-mechanical behavior of identical particles manifests itself most directly through multi-particle interferences and correlations beyond the single-particle level. In particular, second-order spatio-temporal interference provides access to exchange symmetry and quantum statistics even for particles emitted from independent sources. 
Nowadays, higher-order correlations have become a central tool for probing indistinguishability, coherence, and many-particle interference across a wide range of physical platforms~\cite{Brown:1956,Hanbury-Brown:1956,Hong:1987,Glauber:2006,Agne:2017,Menssen:2017,Pleinert:2021,Genovese:2016,Richter:2021,Thekkadath:2022,Roeder:2024,Defienne:2024,Liu:2026,Kiesel:2002,Lougovski:2011,Baym:2014,Kodama:2011,Keramati:2020,Kuwahara:2021,Batelaan:2021,Kuwahara:2021a}. 

Much of this progress has emerged in the context of quantum optics, where photon correlation measurements are routinely employed to characterize nonclassical light and multi-photon interference~\cite{Brown:1956,Hanbury-Brown:1956,Hong:1987,Glauber:2006,Agne:2017,Menssen:2017,Pleinert:2021}, enabling applications in quantum imaging, quantum spectroscopy, and quantum information processing~\cite{Genovese:2016,Richter:2021,Thekkadath:2022,Roeder:2024,Defienne:2024,Liu:2026}. Owing to the bosonic nature of photons and the high degree of control in optical systems, multi-particle interference phenomena have been explored extensively in photonic platforms. Yet, extending these concepts to electrons introduces additional challenges arising from new degrees of freedom, in particular particle interactions and fermionic exchange symmetry.
While bosons exhibit enhanced coincidence probabilities, identical fermions display the Pauli exclusion principle (PEP) as a direct consequence of the antisymmetry of their total wave function.
For electrons, however, exchange-induced antibunching is accompanied by an additional mechanism that suppresses coincidence events: Coulomb repulsion. Since both effects reduce the probability of joint detection, disentangling quantum-statistical correlations from interaction-driven dynamics is inherently nontrivial. This challenge becomes particularly pronounced in free-space geometries, where the long-range nature of the Coulomb interaction remains largely unscreened. As a consequence, to isolate experimentally the clear signature of fermionic exchange symmetry and antibunching in electron Hanbury Brown and Twiss (HBT) experiments has remained difficult~\cite{Kiesel:2002,Lougovski:2011,Baym:2014,Kodama:2011,Keramati:2020,Kuwahara:2021,Batelaan:2021,Kuwahara:2021a}. 
Recent progress in ultrafast electron sources and laser-triggered nanotip emitters has renewed interest in free-space electron interferometry~\cite{Hommelhoff:2006a,Ropers:2007,Barwick:2007,Ehberger:2015,Meier:2018,Keramati:2021,Haindl:2023,Meier:2023,Bruckner:2024,Classen:2023}. 
In particular, spatially coherent and pulsed electron emission enables the preparation of electron pairs with well-defined temporal and spatial properties. This opens the prospect of extending concepts from optical intensity interferometry to charged fermions and of exploiting higher-order correlations of electrons for emerging applications in imaging and spectroscopy~\cite{Genovese:2016,Richter:2021,Thekkadath:2022,Roeder:2024,Defienne:2024}.

In a previous work, we proposed a free-space Hanbury Brown and Twiss geometry employing two independent needle-tip emitters to explore the interplay of Coulomb repulsion and fermionic antibunching~\cite{Classen:2023}. 
In particular, within the semiclassical framework of the study, it was shown that the use of spatially separated emitters allows one to disentangle Coulomb repulsion from fermionic antibunching by encoding both effects differently in the spatial coincidence pattern~\cite{Classen:2023}. 
In contrast to single-tip approaches, the separation between the emitters acts as a tunable knob enabling the control of the relative interplay between the fermionic interference pattern and Coulomb interaction at the source. 
Laser-triggered needle emitters~\cite{Hommelhoff:2006a,Ropers:2007,Barwick:2007,Ehberger:2015,Meier:2018,Keramati:2021,Haindl:2023,Meier:2023,Bruckner:2024} are particularly promising in this context due to their highly controllable emission characteristics, e.g., high spatial coherence and precise temporal operation.

In the present work, we revisit this setup, outlined in Section~\ref{sec:setup}, but now employing for the study a fully quantum-mechanical framework. 
Rather than treating Coulomb interaction classically as in~\cite{Classen:2023}, we formulate the problem as a genuine quantum-mechanical two-body scattering phenomenon of indistinguishable charged fermions in free space (see Section~\ref{sec:framework}).
This allows us to treat exchange statistics and Coulomb interaction on equal footing within a unified quantum-mechanical description.
To this end, we derive in Section~\ref{sec:Model} an analytical model for the two-electron wave function and the associated second-order correlation function in the needle-tip geometry.
In Section~\ref{sec:results}, we discuss the resulting interference patterns and their physical implications. In particular, we demonstrate that the proposed geometry provides a robust platform for disentangling Coulomb interaction and fermionic antibunching in free space.

\section{Needle tips setup}\label{sec:setup}

\begin{figure}[b]
\centering
\includegraphics[width=0.85\columnwidth]{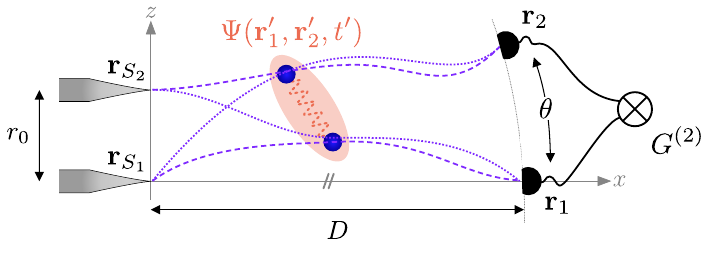}
\caption{Setup of detecting the spatio-temporal correlations of two electrons emitted by two needle tips at $\mathbf{r}_{S_1}$ and $\mathbf{r}_{S_2}$ with inter-slit distance $r_0$. The needle tips emit single electrons, which - after initial acceleration by a bias voltage (not shown) - propagate freely where the state of the two electrons is described by the two-particle wave function $\Psi(\mathbf{r}'_1,\mathbf{r}'_2,t')$. In the far field at $x\approx D$, the electrons are registered by two detectors at $\mathbf{r}_{1}$ and $\mathbf{r}_{2}$ with relative angle $\theta$, allowing to obtain the second-order correlation function $G^{(2)}(\mathbf{r}_{1},\mathbf{r}_{2},t)\propto |\Psi(\mathbf{r}_1,\mathbf{r}_2,t)|^2$.
}
\label{fig:Setup}
\end{figure}

We consider the following setup for the investigation of a Hanbury Brown and Twiss-like experiment with electrons: 
Two independent single-electron sources are placed at positions $\mathbf{r}_{S_1}$ and $\mathbf{r}_{S_2}$ with emitter distance $r_0$ along the $z$-axis as shown in Fig.~\ref{fig:Setup}. 
Each source emits electrons in a precise pulsed manner, e.g., by illuminating needle tips with femtosecond laser pulses. 
In such needle experiments, the electrons are typically accelerated towards the screen by bias voltages $U_\mathrm{b}$ of the order of tens to a few hundred volts \cite{Meier:2023}, where the acceleration can be considered instantaneous~\cite{Classen:2023}.
The electrons can thus be treated non-relativistically with an initial kinetic energy given by $E_{kin}=e \cdot U_\mathrm{b}$. 
In the far field, the resulting correlations of the two electrons are measured at two detectors at positions $\mathbf{r}_{1},\mathbf{r}_{2}$, allowing to obtain the 
second-order spatial correlation function $G^{(2)}(\mathbf{r}_{1},\mathbf{r}_{2})$.  
The latter expresses how likely it is to simultaneously detect an electron at position $\mathbf{r}_{1}$ and a further electron at position $\mathbf{r}_{2}$. 
We assume that the electrons travel without any external influence from the sources to the detectors. A more detailed presentation of the setup can be found in Ref.~\cite{Classen:2023}.

\section{Theoretical framework}\label{sec:framework}

In order to derive $G^{(2)}(\mathbf{r}_{1},\mathbf{r}_{2})$, we formulate the Hanbury Brown and
Twiss-like experiment described above as a quantum-mechanical scattering problem. 
Note that single-particle states do not contribute to the correlation signal, while contributions from three- and higher-particle states can be strongly suppressed in typical experimental realizations using needle tips~\cite{Meier:2023}.
Consequently, we focus exclusively on two-particle states and can thus relate the second-order intensity correlation function to the joint detection probability of two particles at positions $\mathbf{r}_1$ and $\mathbf{r}_2$ (see also Appendix A)
\begin{equation}
    G^{(2)}(\mathbf{r}_1,\mathbf{r}_2) = \mathcal{N} |\Psi(\mathbf{r}_1,\mathbf{r}_2,t)|^2\,.
    \label{eq:G2}
\end{equation}
Here, $\Psi(\mathbf{r}_1,\mathbf{r}_2,t)$ denotes the two-particle wavefunction at the common detection time $t=t_1=t_2$, corresponding to a vanishing detection-time delay $\tau=t_2-t_1=0$, and
$\mathcal{N}$ is an appropriate normalization factor which we will determine in Section~\ref{sec:connection-path-form}.

\subsection{Schrödinger equation of the two-electron problem}\label{sec:Schrödinger}

The time evolution of two electrons of mass $m_e$ and charge $e$
is given by the following two-particle Schrödinger equation
\begin{equation}
    \mathrm{i} \hbar \frac{\partial}{\partial t} \Psi (\mathbf{r}_1,\mathbf{r}_2,t)=
    \left( \frac{\mathbf{\hat{p}}_1^2}{2 m_e}+\frac{\mathbf{\hat{p}}_2^2}{2 m_e}+\frac{e^2}{4 \pi \epsilon_0 \left|\mathbf{r}_1\!-\!\mathbf{r}_2 \right|}\right) \Psi(\mathbf{r}_1,\mathbf{r}_2,t) = \hat{H}  \Psi(\mathbf{r}_1,\mathbf{r}_2,t) \, ,
    \label{SchrödingerEQ_Normal}
\end{equation}
where $\hat{H}$ is the scattering Hamiltonian, comprising the kinetic energy operators of the two particles (with $\mathbf{\hat{p}}_i$ denoting the momentum operator of particle $i$) and their Coulomb interaction.
Similar to the solution of the hydrogen atom~\cite{QuantumMechanics,QuantenmechanikInBildern}, it is convenient to factorize the total wave function 
$$\Psi(\mathbf{r}_1,\mathbf{r}_2,t) \equiv \Psi(\mathbf{R},\mathbf{r},t) =\psi(\mathbf{R},t) \Phi(\mathbf{r},t)$$
into a center-of-mass system (CMS), $\psi(\mathbf{R},t)$ with $\mathbf{R}=(\mathbf{r}_1+\mathbf{r}_2)/2$ and a relative system, $\Phi(\mathbf{r},t)$ with $\mathbf{r}=\mathbf{r}_1-\mathbf{r}_2$, since the potential only depends on the relative distance between the two electrons.
Note that the corresponding relative momentum is defined as
$\mathbf{p}=(\mathbf{p}_1-\mathbf{p}_2)/2$ and the associated relative wave vector by $\mathbf{k}=(\mathbf{k}_1-\mathbf{k}_2)/2$, with wavenumber $k=|\mathbf{k}|$.

\subsection{Spin states and fermionic nature of the electrons}

In the non-relativistic limit considered here, the two-electron wave function can be factorized into a product of a spin part and a spatial part. 
The spin state of two electrons comprises the three symmetric triplet states ($S=1$) and the antisymmetric singlet state ($S=0$).
Considering the fermionic nature of the two indistinguishable electrons, their total wavefunction
must be antisymmetric under exchange of the two particles ($\mathbf{r}_1 \leftrightarrow \mathbf{r}_2$).  
Accordingly, depending on the spin state $S$, 
the spatial wave function takes the form
\begin{equation}
\Psi_S(\mathbf{R},\mathbf{r},t)=\begin{cases}\frac{1}{\sqrt{2}} \left[ \Psi(\mathbf{R},\mathbf{r},t)+\Psi(\mathbf{R},-\mathbf{r},t) \right], & \text{for $S=0$}\,,\\
   \frac{1}{\sqrt{2}} \left[\Psi(\mathbf{R},\mathbf{r},t) -\Psi(\mathbf{R},-\mathbf{r},t) \right], & \text{for $S=1$}\,.
   \end{cases}
   \label{ANTISYM}
\end{equation}
Note that the \mbox{(anti-)}symmetrization of the spatial wavefunction is completely contained in the relative frame.
The separation of the wavefunction into center of mass and relative system is further conserved in time~\cite{QuantumMechanics}. 
We can thus write
\begin{equation}
\Psi_S(\mathbf{R},\mathbf{r},t)=\psi(\mathbf{R},t) \Phi_S(\mathbf{r},t)
\label{eq:PSI_AS_Product}
\end{equation}
with unchanged CMS wavefunction $\psi(\mathbf{R},t)$ and (anti-)symmetrized relative wavefunction 
\begin{equation}
\Phi_S(\mathbf{r},t) =\begin{cases}\frac{1}{\sqrt{2}}\left[\Phi(\mathbf{r},t) + \Phi(-\mathbf{r},t)\right]& \text{for $S=0$}\,,\\
\frac{1}{\sqrt{2}}\left[\Phi(\mathbf{r},t) - \Phi(-\mathbf{r},t)\right]& \text{for $S=1$}\,,\\    
\end{cases} 
\label{eq:phi_AS}
\end{equation}
Inserted into the second-order correlation function in Eq.~\eqref{eq:G2}, we get 
\begin{equation}
    G^{(2)}(\mathbf{r}_1,\mathbf{r}_2,t) = 
    \mathcal{N} \sum_{S,M_S} |\Psi_S(\mathbf{R},\mathbf{r},t)|^2
    =\mathcal{N}|\psi(\mathbf{R},t)|^2\sum_S g_S |\Phi_{S}(\mathbf{r},t)|^2 \, ,
    \label{eq:G2LinkedToWF}
 \end{equation}
where the sum in the first expression extends over all allowed spin states, $S=0,1$ and projections
$M_S=-S,\ldots,S$. Since $\Psi_S(\mathbf{R},\mathbf{r},t)$ is independent of $M_S$,
the sum over the spin projection can be carried out explicitly,
yielding the spin multiplicity $g_S=2S+1$.
Note that the center-of-mass motion will contribute only as an overall envelope, which can be obtained analytically in many cases (e.g., simple plane waves, see Appendix B).
The interesting and nontrivial physics is thus contained in the relative frame on which we will focus in the remainder of the paper.

\subsection{Solution for the relative system}
\label{sec:Solution of the relative system}
In the relative frame, the Schrödinger equation becomes
\begin{equation}
    \mathrm{i} \hbar  \frac{\partial}{\partial t} \Phi (\mathbf{r},t)=\left(-\hbar^2 \frac{\Delta_\mathbf{r}}{2 \mu}+\frac{e^2}{4 \pi \epsilon_0 \left|\mathbf{r}\right|}\right) \Phi (\mathbf{r},t) = \hat{H}_{\mathbf{r}} \Phi (\mathbf{r},t)\, ,
\end{equation}
where $\mu=m_e/2$ is the reduced mass and $\Delta_\mathbf{r}$ the Laplacian in the relative coordinate $\mathbf{r}$.
In analogy to the solution of the hydrogen atom problem, the spatial eigenfunctions of the relative scattering Hamiltonian can be written as~\cite{QuantumMechanics}
\begin{align}
    \Phi_{k,l,m}(\mathbf{r}) &= R_{k,l}(r) Y_{l,m}(\vartheta,\varphi), 
    \label{RelSolution}
\end{align}
where the angular part of the Schrödinger equation is solved by spherical harmonics $Y_{l,m}(\vartheta,\varphi)$ with angular momentum and magnetic quantum numbers $l$ and $m$, respectively, and spatial angles $\vartheta,\varphi$. 
In difference to the hydrogen atom, however, the potential is repulsive and the radial part is solved by the so-called regular Coulomb wave functions~\cite{QuantumMechanics}
\begin{equation}
    R_{k,l}(r)=\frac{A_{k,l}}{r} \mathrm{e}^{\mathrm{i} k r} (k r)^{l+1} F(l+1+\mathrm{i} \eta_k|2(l+1)|-2 \mathrm{i} k r),
    \label{RADIALexact}
\end{equation}
with
\begin{equation}
    F(a|b|z)=\sum_{n=0}^{\infty} \frac{a_n}{b_n}\frac{z^{n}}{n!} \, ,
    \label{HyperGeometricFunction}
\end{equation}
being the confluent hypergeometric function of first kind and $a_n$ ($b_n$) denoting  rising factorials given by $a_n=\Pi^{n-1}_{i=0}(a+i)$ ($b_n$ accordingly). 
Moreover, the prefactor $A_{k,l}$ is given by~\cite{QuantumMechanics}
\begin{equation}
    A_{k,l}=\frac{2^l}{(2l+1)!} \mathrm{e}^{-\frac{1}{2}\pi \eta_k} |\Gamma(l+1+\mathrm{i}\eta_k)|\, ,
\end{equation}
where $\Gamma$ is the Gamma function and $\eta_k$ reads
\begin{equation}
    \eta_k= \frac{e^2}{4\pi \epsilon_0} \frac{\mu}{\hbar^2 k}. 
    \label{ETA}
\end{equation}
Hereby, the time evolution of the spatial eigenfunctions $\Phi_{k,l,m}(\mathbf{r})$ is given by 
\begin{align}
    \Phi_{k,l,m}(\mathbf{r},t) &=  \Phi_{k,l,m}(\mathbf{r})  \exp\left(-\frac{\mathrm{i} E_k t}{\hbar}\right) \, ,
    \label{RelSolution-time}
\end{align}
where $E_k = \hbar^2 k^2/(2\mu)$ is the energy eigenvalue of the continuous spectrum, expressed in terms of the relative wave number $k$.
Knowing the initial state of the two-electron system, $\Phi(\mathbf{r},t=0)$, we can write the general solution in terms of these eigenfunctions of the scattering Hamiltonian via
\begin{equation}
    \Phi(\mathbf{r},t)=\int^\infty_{0}\! \mathrm{d}{k'} k'^2 \sum^{\infty}_{l=0} \sum^l_{m=-l} \zeta_{k',l,m} \Phi_{k',l,m}(\mathbf{r}) \mathrm{e}^{-\mathrm{i}\omega_{k'} t} \, ,
    \label{RelativeFormalSolution}
\end{equation}
with 
\begin{equation}
    \zeta_{k',l,m}=\int_{\Omega_r}\! \Phi(\mathbf{r},t=0) \Phi^*_{k',l,m}(\mathbf{r}) \mathrm{d}\mathbf{r} \,,
    \label{OverlapIntegralRel}
\end{equation}
being the overlap integral of the initial state $\Phi(\mathbf{r},t=0)$ and the spatial eigenfunctions of the relative scattering Hamiltonian given in Eq.~\eqref{RelSolution}. 

Note that Eqs.~\eqref{RelativeFormalSolution} and~\eqref{OverlapIntegralRel} provide the general solution for two distinguishable charged particles. 
For indistinguishable electrons, considered in our case, the state still needs to be \mbox{(anti-)}symmetrized according to Eq.~\eqref{eq:phi_AS}. 

\section{Idealized model for the system of two needle tips}
\label{sec:Model}

In this section, 
we apply the theoretical framework outlined in Section 3 to the specific system of two needle tips.

\subsection{Point-like needle tips.}\label{sec:point-like-tips}

To start with, we assume the tips as point-like single-electron emitters exhibiting identical emission characteristics, resulting in equal spatial mode profiles of the emitted electrons. 
This 
ensures that no which-source information is encoded in the emitted wave packets.
In the case of pulsed operation using sufficiently short pulses, we further assume that the electrons are emitted simultaneously, and we neglect any interaction between electrons from different pulses. 

Under these assumptions, we can model the initial spatial wavefunction of the two electrons as perfectly localized at $t=0$ at the source positions, i.e., 
\begin{equation}
    \Phi_S(\mathbf{r}, t=0)=
    \begin{cases} 
        \frac{1}{\sqrt{2}}\left[\delta(\mathbf{r}-\mathbf{r}_0) 
    + \delta(\mathbf{r}+\mathbf{r}_0)\right], & \text{for $S=0$}\,,\\
        \frac{1}{\sqrt{2}}\left[\delta(\mathbf{r}-\mathbf{r}_0) 
    - \delta(\mathbf{r}+\mathbf{r}_0)\right], & \text{for $S=1$}\,.
   \end{cases}
    \label{Delta_initial}
\end{equation}
where $\mathbf{r}_0$ denotes the vector connecting the two needle tips. 
Note that in Eq.~\eqref{Delta_initial} we incorporated the fermionic nature already in the initial state, where the plus (minus) sign in the expression corresponds to a symmetric (antisymmetric) spatial state. 

Inserting the initial state of Eq.~\eqref{Delta_initial} into the overlap integral of Eq.~\eqref{OverlapIntegralRel}, we find
\begin{equation}
     \zeta_{k,l,m}= 
     \begin{cases} 
        \frac{1}{\sqrt{2}} (1 + (-1)^{l}) \Phi^*_{k,l,m}(\mathbf{r}_0), & \text{for $S=0$}\,,\\
        \frac{1}{\sqrt{2}} (1 - (-1)^{l}) \Phi^*_{k,l,m}(\mathbf{r}_0), & \text{for $S=1$}\,.
   \end{cases}
     \label{Coeff}
\end{equation}
where we applied the property of the delta distribution and $(-1)^{l}$ is due to the parity characteristics of the spherical harmonics. 
Note that the exchange symmetry of the spatial wavefunction determines whether the parity factor $(1\pm(-1)^{l})$ restricts the sum in Eq.~\eqref{RelativeFormalSolution} to even or odd values of $l$. 
Inserting Eq.~\eqref{Coeff} into Eq.~\eqref{RelativeFormalSolution}, the relative wavefunction can be written as
\begin{equation}
    \Phi_S(\mathbf{r},t)_{}= \int \! \mathrm{d}{k'}k'^2\sum^{\infty}_{\substack{l=0}} \Lambda^{S,l} R^*_{k',l}(r_0)R_{k',l}(r)  P_l(\cos \theta/2)\mathrm{e}^{-\mathrm{i}\omega_{k'} t}\, ,
    \label{Delta Inserted}
\end{equation}
with $\theta$ being the angle between the two detectors (see Fig.~\ref{fig:Setup}), related to the polar angle $\vartheta$ of the relative coordinate $\mathbf{r}$ via $\theta/2 = \pi - \vartheta$~\footnote{The latter relation follows geometrically in our system from the far-field configuration: the relative coordinate $\mathbf r$ points along the chord connecting both detector positions on the detection sphere, such that its polar angle $\vartheta$ is related to half the detector relative angle $\theta$.}. Note that in deriving Eq.~\eqref{Delta Inserted}, we exploited the addition theorem of angular harmonics, i.e., 
\begin{equation}
\sum_{m=-l}^{m=l}Y^*_{l,m}(\vartheta_0,\varphi_0) Y_{l,m}(\vartheta,\varphi) = \frac{2l+1}{4\pi} P_{l}(\cos(\pi-\vartheta))= \frac{2l+1}{4\pi} P_l(\cos \theta/2)
\end{equation}
with $P_l(x)$ being the  Legendre polynomials; we further introduced the prefactor
\begin{equation}\label{eq:Lamda-S-l}
    \Lambda_{S,l}=\begin{dcases}
        \frac{1+(-1)^{l}}{\sqrt{2}}\frac{(2l\!+\!1)}{ 4 \pi} \, , \quad  \mathrm{for}\, S=0 \, ,\\
        \frac{1-(-1)^{l}}{\sqrt{2}}\frac{(2l\!+\!1)}{ 4 \pi} \, , \quad \mathrm{for}\, S=1 \, 
    \end{dcases}
\end{equation}
that incorporates the spin-dependency.

\subsection{Monochromatic emission.}

We next assume that the emitted electrons are monochromatic, i.e., $|\mathbf{k}_1|=|\mathbf{k}_2|=\kappa$. Strictly speaking, perfectly localized initial states and monochromatic emission are mutually incompatible as a consequence of Heisenberg's uncertainty principle. Nevertheless, both can be approximated by combining strongly localized emission from very sharp needle tips with post-selection of detection events based on ultrashort electron time-of-flight measurements, which selects a narrow energy distribution. 
Note that while the laboratory-frame wavenumber $\kappa$ is now fixed, the relative wavenumber $k=|\mathbf{k}_1-\mathbf{k_2}|/2$ can still vary with the detector positions. For monochromatic emission and the detector setup of Fig.~\ref{fig:Setup}, we can relate the two via $k=\kappa\sin(\theta/2)$ with $\theta$ being the relative angle between the detectors.

For a given $\theta$, monochromatic electron emission is imposed by introducing the filter function $N_F=1/{k'}^2\delta(k-k')$ into Eq.~\eqref{Delta Inserted}, in which case the integration over $k'$ in Eq.~\eqref{Delta Inserted} leads to
\begin{equation}
    \Phi_{S,k}(\mathbf{r},t)=  \sum^{\infty}_{l=0} \Lambda_{S,l} R^*_{k,l}(r_0)R_{k,l}(r)  P_l(\cos(\theta /2)) \mathrm{e}^{-\mathrm{i}\omega_k t}
    \label{SimMod}
\end{equation}
where the subscript $k$ indicates monochromatic emission.

\subsection{Far-field approximation.}

The general solutions of the relative Schrödinger equation, the radial functions $R_{k,l}$ of Eq.~\eqref{RADIALexact}, contain incoming and outgoing contributions~\cite{QuantumMechanics}.
These radial functions enter two times in $\Phi_{S,k}$ of Eq.~\eqref{SimMod}: in the projection of the initial state and in the subsequent evolution. 
It is important to distinguish these two roles. 
For the expansion of the initial state, a complete set of eigenfunctions is required, which is naturally provided by the regular Coulomb solutions $R_{k,l}(r_0)$ that are well behaved at the origin. 
The subsequent evolution, however, corresponds in our setup to particles emitted from localized sources and detected in the far field, implying outgoing scattering boundary conditions. 
Accordingly, we will use for the second radial function $R_{k,l}(r)$ only the outgoing solution $R^{(out)}_{k,l}(r)$~\cite{QuantumMechanics}.
The product of the two radial functions in Eq.~\eqref{SimMod} can then be approximated by

\begin{equation}
   R_{k,l}^*(r_0)R^{(out)}_{k,l}(r) \approx R^*_{k,l}(r_0)\frac{1}{r} \exp\left[ \mathrm{i}\left( kr-\eta \ln(2 k r) - \frac{l \pi}{2} +\delta_l\right)\right],
   \label{farfield}
\end{equation}
where the so-called Coulomb scattering phase is given by 
\begin{equation}
    \delta_l=\frac{1}{2 \mathrm{i}} \log\left(\frac{\Gamma(l+1+\mathrm{i}\eta)}{\Gamma(l+1-\mathrm{i}\eta)}\right).
    \label{Scatterphase}
\end{equation}

\subsection{Resulting second-order correlation function.}\label{sec:ResultingWaveFunction}

Combining Eqs.~\eqref{SimMod},~\eqref{farfield}, and~\eqref{Scatterphase}, the resulting wave function within the outlined approximations is given by
\begin{equation}
    \Phi_{S,k}(\mathbf{r},t)= \frac{1}{r}\exp\left[\mathrm{i}(kr-\omega_k t)\right] \exp\left[-\mathrm{i}\eta\ln (2kr)\right] \sum^{\infty}_{l=0} \Lambda_{S,l} R^*_{k,l}(r_0)   e^{-\mathrm{i}l\pi/2} e^{\mathrm{i}\delta_l}  P_l(\cos(\theta /2)) \, .
    \label{psi_Workhorse}
\end{equation}
The second-order correlation function of Eq.~\eqref{eq:G2LinkedToWF} can thus be written as
\begin{eqnarray}
    G^{(2)}(\theta) 
    &=\mathcal{N} \sum_S g_S\left|\frac{1}{r}\sum^{\infty}_{l=0} \Lambda_{S,l} \mathrm{i}^l R_{k,l}(r_0) \mathrm{e}^{-\mathrm{i}\delta_l} P_l(\cos  \theta/2)\right|^2 \, .
    \label{Workhorse}
\end{eqnarray}
%
where $G^{(2)}(\theta)$ is evaluated as a function of the relative angle $\theta$, with $\sin(\theta/2)= |\mathbf{r}_1-\mathbf{r_2}|/(2D)$ (see Fig.~\ref{fig:Setup}).
From Eq.~\eqref{Workhorse} we can see that the overall structure of $G^{(2)}(\theta)$ is governed by the tip separation $r_0$, the detector distance $D$, the electron energy (through the laboratory-frame wavenumber $\kappa$ and the corresponding relative wavenumber $k$), and the two-electron spin $S$.
Note that
$G^{(2)}(\theta)$ is time-independent 
as the time evolution contributes only with a global phase in Eq.~\eqref{psi_Workhorse}.
The so far disregarded center-of-mass system likewise contributes only with a global phase and therefore does not modify $G^{(2)}$ (see Appendix B).
%

Equation~\eqref{Workhorse} will serve as the main  expression for the remainder of this paper. 
It describes the system within an idealized model of point-like needle tips, monochromatic emission, and far-field detection.
Despite these idealizations, the model still captures the essential physics, as demonstrated below.

\subsection{Connection to the quantum path formalism in the uncharged case}\label{sec:connection-path-form}

\begin{figure}[b]
\centering\includegraphics[width=0.95\columnwidth]{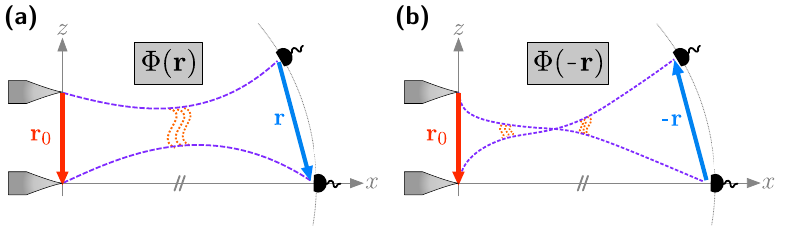}
\caption{Illustration of the quantum paths corresponding to $\Phi(\mathbf{r})$ and $\Phi(-\mathbf{r})$. In both figures, $\mathbf{r}_0$ corresponds to the initial separation of the two electrons emitted at the two needle tips. The dashed violet lines represents the two possible two-electron paths, and the orange dotted lines illustrate the Coulomb interaction.
\textbf{(a)} $\Phi(\mathbf{r})$ corresponds to the parallel path in the quantum path picture, where the electron from the upper (lower) needle tip is registered at the upper (lower) detector. \textbf{(b)} $\Phi(-\mathbf{r})$ corresponds to the diagonal path, where the electron from the upper (lower) needle tip is registered at the lower (upper) detector; here, the electrons have to `cross' each other. 
}
\label{Pfade}
\end{figure}

In~\cite{Classen:2023}, two-electron interference from two tips was derived in a semiclassical model. In particular, the interference pattern in the case of neutral fermions was derived using a quantum path framework.
The latter predicts the interference of two indistinguishable two-particle paths, a `parallel' one and a `crossed' one as shown in Fig.~\ref{Pfade} (a) and (b), respectively. 
In the quantum-mechanical wavefunction formalism discussed in this paper, these two distinct two-electron paths correspond to $\Phi(\mathbf{r})$ and $\Phi(\mathbf{-r})$ [see Fig.~\ref{Pfade} and Eq.~\eqref{eq:phi_AS}].
Due to their indistinguishability, the two two-electron paths have to be superposed coherently, i.e., $\Phi(\mathbf{r}) \pm \Phi(\mathbf{-r})$.
In the quantum-mechanical wavefunction formalism, on the other hand, the above superposition corresponds to the \mbox{(anti-)}symmetrization of the two-particle wavefunction $\Phi_{S}(\mathbf{r})\propto \Phi(\mathbf{r}) \pm \Phi(\mathbf{-r})$.
In this way, the quantum path formalism emerges naturally from (anti-)symmetrization.
%
%

%
Quantitatively, in the absence of Coulomb interaction, i.e., in the limit $q\rightarrow 0$, $\eta_k$ of Eq.~\eqref{ETA} as well as the Coulomb scattering phase $\delta_l$ of Eq.~\eqref{Scatterphase} vanish and 
the Coulomb radial functions reduce to the free-particle radial solutions, i.e., spherical Bessel functions~\cite{QuantumMechanics}
\begin{equation}
    R_{k,l}(r_0) \overset{\eta\rightarrow 0}{\longrightarrow}k j_{k,l}(r_0) .
    \label{ChargelessRadial}
\end{equation}  
The second-order correlation function $G^{(2)}$ of Eq.~\eqref{Workhorse} then takes the form
\begin{align}\label{eq:EtaToZero}
    G^{(2)}(\theta) &=\mathcal{N}\sum_S g_S\left|\frac{1}{r}\sum^{\infty}_{l=0} \Lambda_{S,l} \mathrm{i}^l j_l(k,r_0) P_l(\cos(\theta /2))\right|^2 
     \\
    &=\mathcal{N}\sum_S g_S\left(\frac{k}{4 \pi r}\right)^2\frac{1}{2}\left| \mathrm{e}^{\mathrm{i} \mathbf{k}\cdot (\mathbf{r}+\mathbf{r}_0)} +(-1) ^S\mathrm{e}^{\mathrm{i} \mathbf{k}\cdot(\mathbf{r}-\mathbf{r}_0)}\right|^2\notag \\
    &=\mathcal{N}\sum_S g_S\left(\frac{k}{4 \pi r}\right)^2\left[1+(-1) ^S\cos(\kappa r_0 \sin{(\theta)} )\right]\notag
\end{align}
where from the first to the second line we expand the parity factor $1\pm (-1)^l$ from $\Lambda_{S,l}$ (see Eq.~\ref{eq:Lamda-S-l}) to obtain two sums, each corresponding to the standard partial-wave expansion of a plane wave in spherical coordinates. 
Moreover, from the second to the third line, we identify $2\mathbf{k} \cdot \mathbf{r_0}=  2\kappa r_0 \sin{(\theta/2)} \cos{(\theta/2)} = \kappa r_0 \sin{(\theta)}$.
Finally, choosing $\mathcal{N}=(4 \pi r/k)^2=(8 \pi D/\kappa)^2$,
we obtain in Eq.~\eqref{eq:EtaToZero} the same result as in the quantum path framework discussed in Ref.~\cite{Classen:2023}.

\section{Results and discussion}\label{sec:results}

To display the interference pattern $G^{(2)}$ of two scattering electrons predicted by the idealized model outlined above, we solve Eq.~\eqref{Workhorse} numerically using Python.

\subsection{Benchmark in the noninteracting limit}

\begin{figure}
\centering\includegraphics[width=0.75\columnwidth]{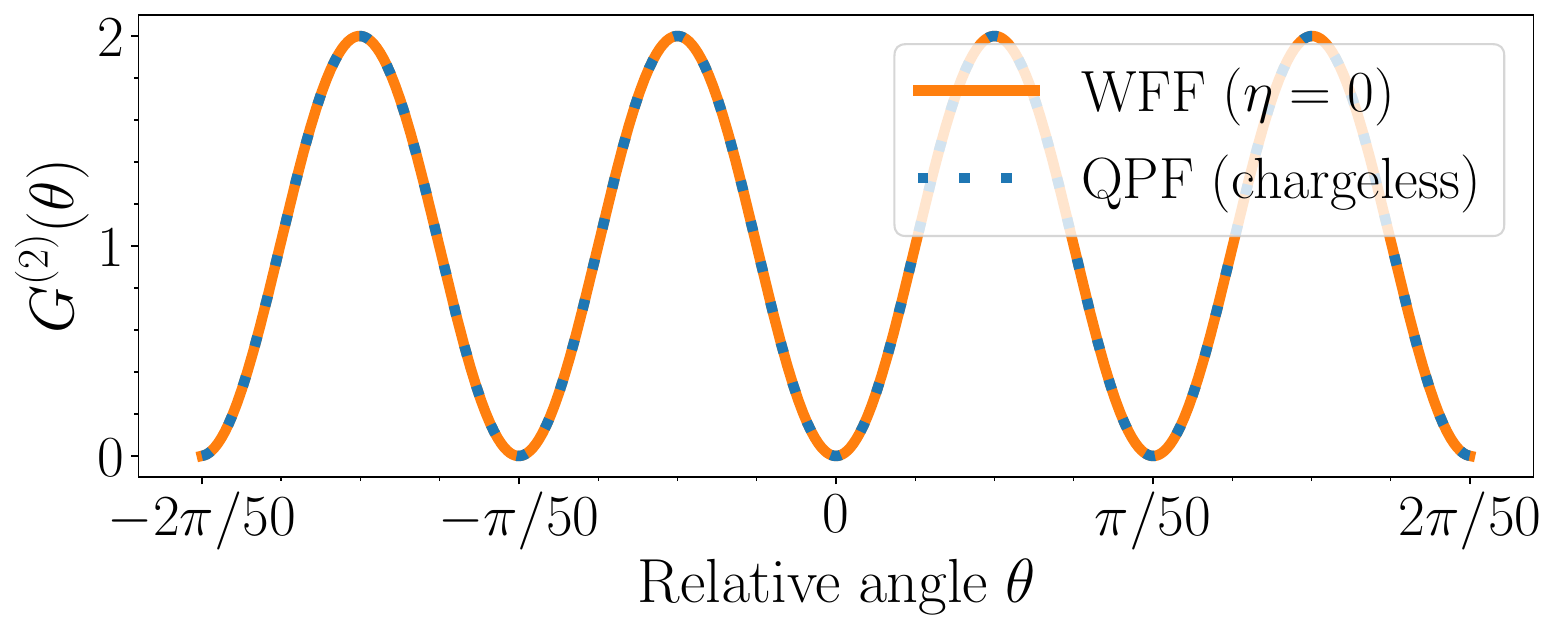}
\caption{
Comparison of the second-order correlation function $G^{(2)}$ for $\kappa=\SI{1e11}{\per\metre}$ and $r_0=\SI{1}{\nm}$,
obtained in the noninteracting regime $\eta=0$ via the numerically evaluated  wavefunction formalism (WFF) (orange solid line) and via the analytical quantum path formalism (QPF) of~\cite{Classen:2023} (blue dotted line). 
Both approaches yield the same spatial interference pattern demonstrating that the result derived via QPF is recovered by the WFF approach in the noninteracting limit.
}
\label{fig:WFF-vs-QPF}
\end{figure}

As a consistency check of the numerical implementation, we compare the results obtained for $G^{(2)}$ in the noninteracting limit $q \rightarrow 0$ with the analytical predictions of the quantum path formalism discussed in~\cite{Classen:2023}. 
To this aim, we use the Coulomb-scattering solution and explicitly evaluate it for $q = 0$. 
In this limit, the numerical solution reproduces the expected interference pattern of two indistinguishable quantum paths (see Fig.~\ref{fig:WFF-vs-QPF}). This benchmark provides an important validation of both the numerical implementation and the employed partial-wave expansion.

\subsection{Coulomb effect in two-electron interference}

\begin{figure}
\centering
\includegraphics[width=0.75\columnwidth]{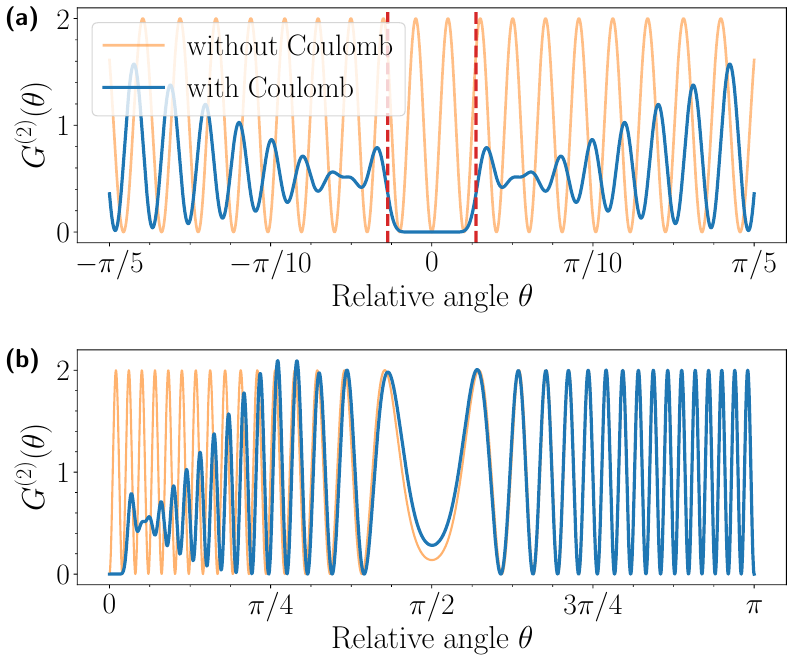}
\caption{Spatial interference pattern $G^{(2)}$ for two electrons with parallel spin ($S=1$, $M_S=1$), with Coulomb interaction (blue) and for reference without Coulomb interaction (orange). 
(a) The interference pattern for small relative angles $\theta$ displays a strong Coulomb suppression; here, the vertical red dashed line corresponds to the classical estimation of the Coulomb dip derived in~\cite{Classen:2023}.
(b) Interference pattern from $\theta=0$ to $\theta=\pi$. At $\theta=0$, the electrons travel in the same direction with strong Coulomb suppression as shown in (a);
at $\theta=\pi$, in contrast, the electrons travel in opposite directions, where Coulomb interaction becomes negligible. Note that the frequency change around $\pi/2$ is a pure geometrical effect.
}
\label{FullPLOT}   
\end{figure}

We now turn to the two-electron problem including Coulomb interaction and investigate how the latter modifies the second-order correlation function $G^{(2)}$.
For clarity, we first restrict the discussion to a single spatially antisymmetric wavefunction, e.g., the triplet state $S=1$, $M_S=1$, before generalizing the result to arbitrary spin configurations.
For this state, Fig.~\ref{FullPLOT} shows the resulting second-order correlation function for the parameters $\kappa=\SI{1e11}{\per\metre}$ and $r_0=\SI{1}{\nm}$. While the dependence on these parameters will be analyzed in more detail below, we first discuss the characteristic features of the obtained interference pattern.

The pattern can be divided into three characteristic regions. In the central region ($\theta \approx 0$), the coincidence signal is strongly suppressed due to Coulomb repulsion. Surrounding this central dip, a transition regime emerges in which interference fringes reappear but remain substantially modified by Coulomb interaction. At larger relative angles $\theta$, the Coulomb contribution finally becomes negligible and the pattern approaches the noninteracting fermionic reference pattern shown in orange in Fig.~\ref{FullPLOT}.
The classical estimate of the Coulomb-dominated region obtained in Ref.~\cite{Classen:2023} is indicated by the red dashed lines in Fig.~\ref{FullPLOT}(a). While this estimate captures the characteristic size of the suppression region remarkably well, the full quantum-mechanical treatment reveals additional features absent in the semiclassical model.
Most notably, an extended Coulomb-dominated region emerges where the coincidence signal is nearly complete suppressed.
Moreover, in the transition regime, the interference fringes exhibit both an offset, which reduces the visibility, and a phase shift relative to the noninteracting fermionic reference pattern.  
Since the phase shift is not constant, with a maximal shift close to the Coulomb dip, the interplay between the nonuniform phase shift and the offset distorts the fringe pattern in the transition regime, resulting in the near suppression of an entire interference fringe.
Finally, with further increasing relative angle $\theta$, the phase shift and the offset become weaker  eventually vanishing in the large-angle limit. 
Thus, for sufficiently large relative angles, both patterns coincide, indicating that Coulomb interaction becomes negligible in this regime; in particular, in the extreme case $\theta\approx\pi$, the electrons propagate approximately in opposite directions such that the Coulomb interaction mainly leads to a mutual acceleration rather than a substantial distortion of the interference structure.
Most importantly, the spatial oscillation frequency remains essentially unchanged relative to the noninteracting fermionic reference pattern outside of the central dip. 

\subsection{Spin dependence of the interference pattern}

\begin{figure}
\centering
\includegraphics[width=0.75\columnwidth]{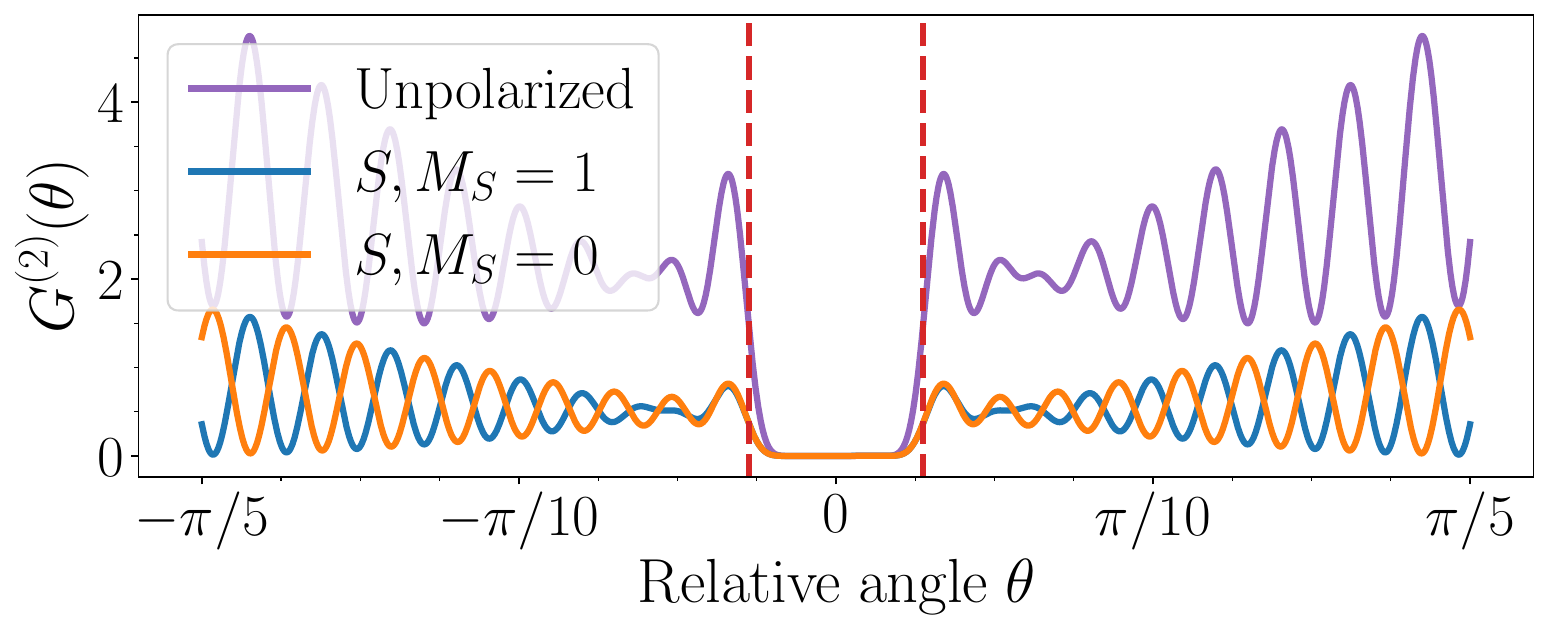}
\caption{Comparison between the spatial interference patterns $G^{(2)}$ for two electrons with parallel spins, i.e., $S=1$, $M_S=1$ (blue solid line), two electrons with antiparallel spins, i.e., $S=0$, $M_S=0$, mimicking 'bosonic' electrons (orange solid line), and two electrons with unpolarized spins (purple solid line). The vertical red dashed line corresponds to the classical estimation of the Coulomb dip as discussed in~\cite{Classen:2023}. 
}
\label{Spin-Plot}   
\end{figure}

In general, spin configurations different from $S=1$, $M_S=1$ also contribute to the interference signal. 
The resulting second-order correlation function 
adding all spin configurations
is shown in Fig.~\ref{Spin-Plot}.

As expected, the central dip is unaffected by the spin configuration, demonstrating that Coulomb repulsion is the dominating contribution here. 
In contrast, the visibility and phase of the interference fringes depends strongly on the spin state. Except next to the Coulomb dip, the spatially symmetric and antisymmetric wavefunctions exhibit a relative phase shift of $\pi$, directly reflecting the underlying exchange symmetry. Consequently, antisymmetric spin states give rise to a spatially symmetric interference pattern analogous to the bosonic case, whereas symmetric spin states lead to the fermionic antibunching pattern discussed in the previous section.

For unpolarized electrons, the observed interference signal results from an incoherent sum over all possible spin configurations, i.e., the three triplet states and the singlet state. Since the triplet states are associated with spatially antisymmetric wavefunctions, their contribution dominates the resulting pattern. 
As a consequence, the overall visibility of the interference fringes is reduced, while the Coulomb-induced suppression remains unchanged. 
Note that the reduced visibility for unpolarized electrons is consistent with the quantum path interpretation of additional spin-dependent exchange paths discussed in~\cite{Classen:2023}.

\subsection{Experimental parameter regimes}

While the wave number value $\kappa=\SI{1e11}{\per\metre}$ considered in the previous examples corresponds to experimentally realistic values, 
the tip separation $r_0=\SI{1}{\nm}$ was intentionally  chosen to highlight the Coulomb-induced modifications of free-space HBT interference of electrons. 
In realistic experimental implementations, however, achievable tip separations are expected to be much larger, on the order of $50-100 \, \rm{nm}$~\cite{Bruckner:2024}.

\begin{figure}
\centering
\includegraphics[width=0.7\columnwidth]{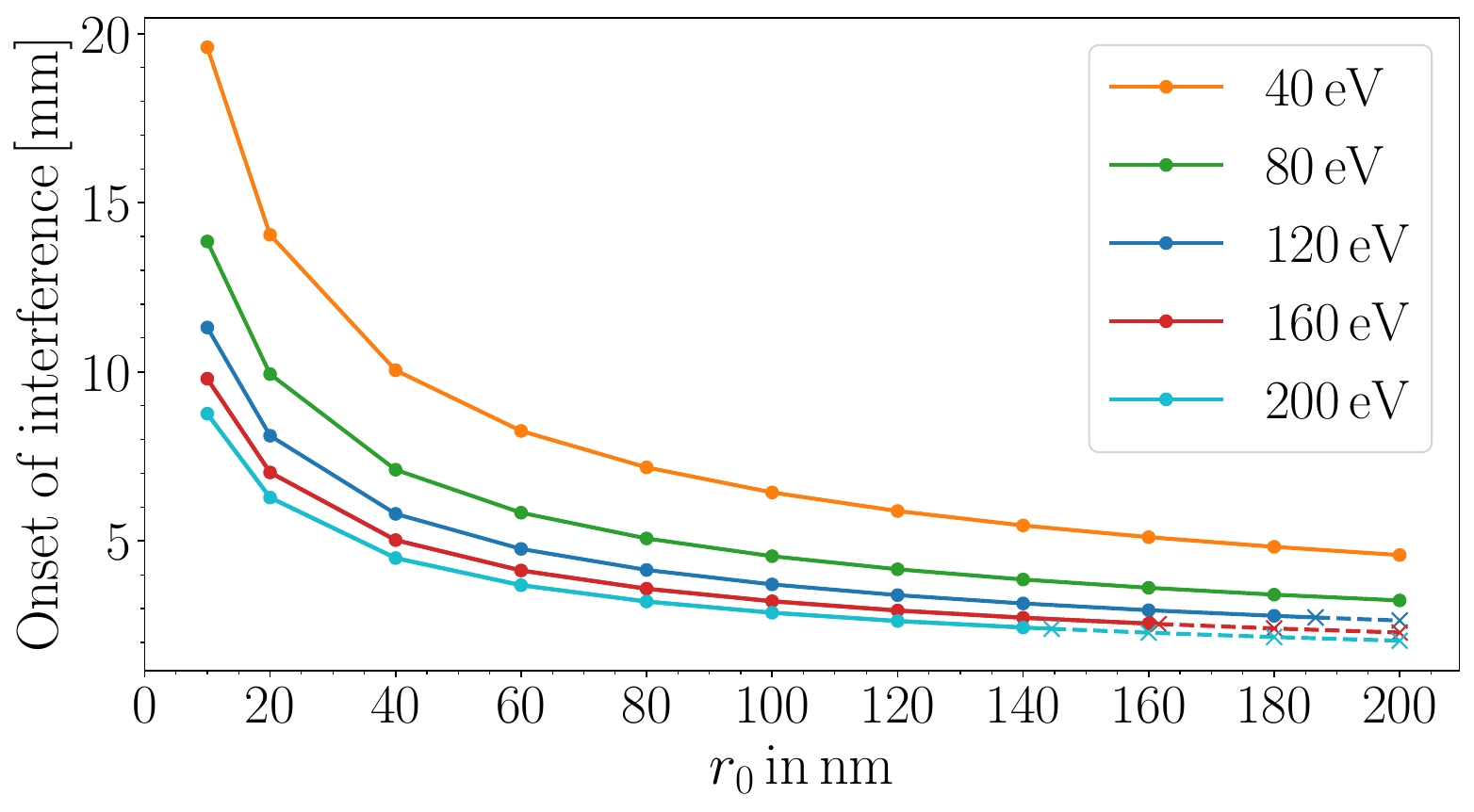}
\caption{
Onset of the oscillatory interference pattern as a function of the tip separation $r_0$ for different bias voltages $U_b=40\,\rm{eV},\dots,200\,\rm{eV}$. 
The onset of the oscillatory regime provides a measure of the width of the Coulomb-dominated region. 
Increasing the bias voltage corresponds to augmenting the initial kinetic energy of the electrons and thus the wave number $\kappa$. 
Here, the tip-to-detector distance was chosen to be $D=\SI{25}{\cm}$~\cite{Meier:2023}.
}
\label{Plot-Parameter-Regime}   
\end{figure}

While within the present model the spatial oscillation frequency scales linearly with $\kappa$ and $r_0$, 
the dependence of the width of the Coulomb-dominated suppression region (quantified by the onset of the oscillatory interference pattern) on $\kappa$ (determined by the bias voltages $U_b$) and on $r_0$
is illustrated in
Figure~\ref{Plot-Parameter-Regime}.
Increasing either parameter, $\kappa$ and $r_0$, reduces the influence of Coulomb repulsion: 
larger tip separations $r_0$ weaken the interaction already at the source, while higher kinetic energies (corresponding to higher $\kappa$ values) shorten the interaction time during propagation. 
Consequently, the onset of the oscillatory interference regime shifts towards smaller detector separations and the pattern approaches more rapidly the noninteracting fermionic limit.

While larger values of $\kappa$ and $r_0$ suppress Coulomb-induced distortions, at the same time, they also lead to increasingly fine interference fringes, eventually exceeding the spatial resolution of currently available detectors with a fringe width resolution of, e.g., $150\,\rm{\mu m}$~\cite{Meier:2023}. 
The latter parameter regime occurs for $r_0 \geq 145 \rm{nm}$ and $U_{\rm{bias}} \geq 120 \rm{eV}$, marked by crosses and dashed lines in Fig.~\ref{Plot-Parameter-Regime}.
Realistic experimental realizations thus require a compromise between minimizing Coulomb effects and maintaining experimentally resolvable interference fringes.

\subsection{Range of validity of the model}

While the present model captures the essential physics of two-electron interference in the presence of Coulomb interaction in a full quantum-mechanical treatment, several simplifying assumptions limit its quantitative applicability to realistic experimental conditions.
First, the electron sources are assumed to be point-like and to emit identical spatial modes. In a realistic setup, the finite spatial extent of the emission region inside the needle tips leads to an effective averaging over different emission positions, which is expected to reduce the visibility of the interference fringes. 
Second, the present treatment assumes simultaneous electron emission. For finite temporal delays between the emission events, Coulomb interaction additionally induces a longitudinal momentum modification between the electrons: the leading electron is accelerated, whereas the trailing electron is decelerated. 
Third, the electrons are assumed to be monochromatic. By contrast, a finite longitudinal momentum distribution would introduce additional averaging over different wave numbers thus further reducing the fringe contrast. In this case, the temporal evolution of the wave packets becomes nontrivial and has to be taken explicitly into account.
Moreover, the finite mass of the electrons results in wave-packet dispersion during propagation.
Such effects are expected to modify both the temporal and spatial correlation structure and will be investigated in future work.

\section{Conclusion}
\label{sec:conclusion}

In this paper, we employed a quantum-mechanical framework in order to derive the two-electron interference pattern resulting from two interacting electrons emitted by two needle tips in the presence of Coulomb interaction. It reproduces the fermionic interference pattern predicted by the quantum path formalism in the noninteracting limit and naturally generalizes the description to interacting electrons. 
Compared to the semiclassical model discussed in Ref.~\cite{Classen:2023}, the quantum-mechanical treatment predicts three additional features: an extended Coulomb-dominated region with nearly complete suppression of the coincidence signal, a Coulomb-induced  offset, and a gradual phase shift of the interference fringes relative to the noninteracting reference pattern; the exact nature of this phase shift is an open question and will be explored in a later study.
\noindent At the same time, the spatial oscillation period of the interference fringes remains essentially governed by the fermionic exchange symmetry and thus agrees with the prediction of the quantum path formalism. In particular, the emergence of visible interference fringes beyond the Coulomb-dominated region identifies experimentally accessible regimes in which Coulomb interaction and fermionic antibunching are clearly disentangled.

We close by noting that the presented solution closely resembles the quantum-mechanical treatment of the hydrogen atom, reflecting the formal analogy between the two problems. While this approach, using spherical coordinates, provides valuable physical insight, it is less efficient for the numerical computation of the interference patterns. A more direct albeit less intuitive formulation can be obtained using parabolic coordinates~\cite{QuantumMechanics}, what significantly reduces the computational effort. This approach will be particularly useful for future extensions of the model, including more realistic initial states such as Gaussian wave packets in real and momentum space as well as different emission times, where the temporal evolution becomes nontrivial.

\ack

This work was funded by the Deutsche Forschungsgemeinschaft (DFG, German Research Foundation) – Projektnummer 581991507.
A.C., M.-O.P., and J.v.Z. acknowledge funding by the Erlangen Graduate School in Advanced Optical Technologies (SAOT) by the Bavarian State Ministry for Science and Art.

\setcounter{section}{0}
\renewcommand{\thesection}{\Alph{section}}

\section{From field operators to Coulomb scattering}
\label{sec:FromFieldToCoulomb}
Equation~\eqref{eq:G2} evaluates the second-order correlation function in first quantization, by identifying $G^{(2)}$ with the joint probability $|\Psi(\mb{r}_1,\mb{r}_2,t)|^2$. For completeness, we show in this appendix that the same expression follows from a treatment using second quantization where $G^{(2)}$ is defined through fermionic field operators evaluated at the detector positions 
~\cite{Classen:2023}. Beyond confirming Eq.~\eqref{eq:G2}, the second quantization route makes two features more explicit that remain implicit in the wavefunction picture: the origin of the exchange symmetry in the anti-commutation relations of the field operators and the statistical weights with which the singlet and triplet channels enter the unpolarized signal.

\subsection{Field operators and correlation function}
Let $\hat{\Psi}_s(\mb{r})$ annihilate an electron of spin projection $s$ at position $\mb{r}$, satisfying the fermionic anti-commutator relations
\begin{equation}
\{\hat{\Psi}_s(\mathbf{r}_a),\hat{\Psi}^\dagger_{s'}(\mathbf{r}_b)\}_{+}=\delta_{s,s'}\delta(\mathbf{r}_a-\mathbf{r}_b),  \;\;\;\; \{\hat{\Psi}_s(\mathbf{r}_a),\hat{\Psi}_{s'}(\mathbf{r}_b)\}_{+}=0  
\end{equation}
as well as $\hat{\Psi}_i(\mb{r}_j)\ket{0}=0$. In the Heisenberg picture, using the two-body Hamiltonian $\hat{H}$ of Eq.~\eqref{SchrödingerEQ_Normal}, the time evolution is given by
\begin{equation}
    \hat{\Psi}_s(\mathbf{r}_i,t)= \mathrm{e}^{\mathrm{i} \frac{\hat{H}t}{\hbar}} \hat{\Psi}_{s}(\mb{r}_i)\,\mathrm{e}^{-\mathrm{i} \frac{\hat{H}t}{\hbar}}.
    \label{Fieldoperator2}
\end{equation}
In contrast to the quantum-path treatment of \cite{Classen:2023}, we do not decompose the field into contributions originating from the two sources. The field operators are rather evaluated directly at the detector positions, and the relative phase of the two quantum paths enters later, through the initial state, rather than through the operators.
The second-order correlation function is the normally ordered expectation value~\cite{Glauber:2006}
\begin{equation}
G^{(2)}(\mb{r}_1,\mb{r}_2)=\sum_{s,s'} G^{(2)}_{s,s'}(\mb{r}_1,\mb{r}_2)=\sum_{s,s'}\left\langle \hat{\Psi}_s^\dagger(\mb{r}_1,t)\hat{\Psi}_{s'}^\dagger(\mb{r}_2,t) \hat{\Psi}_{s'}(\mb{r}_2,t)\hat{\Psi}_s(\mb{r}_1,t)  \right\rangle_\rho \, .\label{eq:G2_sum_Heisenberg2}
\end{equation}
where the sum runs over the spin projections of the two detected electrons and $\langle\,\cdot\,\rangle_\rho=\mathrm{Tr}(\rho\;\cdot\,)$, with $\rho$ the initial state (density matrix) of the electrons. Since the detectors are not spin resolving, all $(s,s')$ contribute incoherently. 
\noindent Inserting the field operators Eq.~\eqref{Fieldoperator2} into Eq.~\eqref{eq:G2_sum_Heisenberg2}, the intermediate time evolution operators between the field operators cancel pairwise, and we obtain 
\begin{equation}
 G^{(2)}(\mb{r}_1,\mb{r}_2)=\sum_{s,s'}G^{(2)}_{s,s'}(\mb{r}_1,\mb{r}_2)= \sum_{s,s'}\left\langle \mathrm{e}^{\mathrm{i} \frac{\hat{H}t}{\hbar}} \hat{\Psi}^\dagger_{s}(\mb{r}_1) \hat{\Psi}_{s'}^\dagger(\mb{r}_2) \hat{\Psi}_{s'}(\mb{r}_2)
  \hat{\Psi}_{s}(\mb{r}_1)  \mathrm{e}^{-\mathrm{i} \frac{\hat{H}t}{\hbar}}\right\rangle_\rho \, .\label{eq:G2_SUM2}  
\end{equation}
For a pure state $\rho=\ket{\Psi_0}\!\!\bra{\Psi_0}$ we obtain 
\begin{equation}
 G^{(2)}_{s,s'}(\mb{r}_1,\mb{r}_2)= \bra{\Psi_0}\mathrm{e}^{\mathrm{i} \frac{\hat{H}t}{\hbar}} \hat{\Psi}^\dagger_{s}(\mb{r}_1) \hat{\Psi}_{s'}^\dagger(\mb{r}_2) \hat{\Psi}_{s'}(\mb{r}_2)
  \hat{\Psi}_{s}(\mb{r}_1) \mathrm{e}^{-\mathrm{i} \frac{\hat{H}t}{\hbar}} \,\ket{\Psi_0} .\label{gss}  
\end{equation}
Equation~\eqref{gss} provides the equivalent description within the Schrödinger picture: the operators are taken at $t=0$ and the time dependence is carried by the state $\mathrm{e}^{-\mathrm{i} \frac{\hat{H}t}{\hbar}} \,\ket{\Psi_0} = \ket{\Psi(t)}$. 
Recall that, as in the main text, we consider equal-time correlation functions and $t$ denotes the common time evolution of the two-particle state.

\subsection{Separation of the Hamiltonian and choice of the basis}
As in Subsection~\ref{sec:Schrödinger}, the two-particle Hamiltonian separates into
\begin{equation}
    \hat{H}=\hat{H}_{CMS}+\hat{H}_{\mb{r}}\, .
    \label{Separation2}
\end{equation}
The two terms commute and act on different coordinates, so the eigenfunctions factorize,
\begin{equation}
    \ket{\Psi}=\ket{\psi}^{{CMS}}\otimes \ket{\Phi}^{rel}=\ket{\psi} \ket{\Phi}
\end{equation}
Hereby, the Coulomb interaction is contained entirely in $\hat{H}_{\mb{r}}$, whereas $\hat{H}_\mathrm{CMS}$ is a free particle Hamiltonian of mass $M=2m_e$.
\noindent We can then expand the state into the respective eigenstates of the Hamiltonian. For the CMS system we get plane waves as eigenfunctions, whereas for the relative system we get Coulomb wavefunctions as introduced in Subsection~\ref{sec:Solution of the relative system}.

\subsection{Time evolution of the initial state}

The time evolution of the state $\ket{\Psi_0}$ can thus be written as
\begin{equation}
    \mathrm{e}^{-\mathrm{i} \frac{\hat{H}t}{\hbar}} \,\ket{\Psi_0}=\mathrm{e}^{-\mathrm{i} \frac{\hat{H}_{CMS}t}{\hbar}}\ket{\psi_0} \otimes \mathrm{e}^{-\mathrm{i} \frac{\hat{H}_{\mb{r}}t}{\hbar}}\ket{\Phi_0}
\end{equation}
In the following, we will consider the center of mass and the relative frame separately. We start with the relative frame, for which we obtain
\begin{equation}
   \mathrm{e}^{-\mathrm{i} \frac{\hat{H}_{\mb{r}} t}{\hbar}} \,\ket{\Phi_0}=\mathrm{e}^{-\mathrm{i} \frac{\hat{H}_rt}{\hbar}} \int \mathrm{d}k' k'^2\sum_{l,m}\ket{\Phi_{k',l,m}}\!\!\braket{\Phi_{k',l,m}|\psi_0}= \int \mathrm{d}k' k'^2\sum_{l,m}\zeta_{k',l,m}  \mathrm{e}^{-\mathrm{i} \frac{E_{k'}t}{\hbar}}\ket{\Phi_{k',l,m}} 
   \label{eq:InsertCompleteness}
\end{equation}
where $\zeta_{k,l,m}=\braket{\Phi_{k,l,m}|
\Phi_0}$ denotes the expansion coefficient [see Eqs.~\eqref{OverlapIntegralRel} and~\eqref{Coeff}]. Correspondingly, in the CMS system we obtain
\begin{equation}
    \mathrm{e}^{-\mathrm{i} \frac{\hat{H}_C t}{\hbar}} \ket{\psi_0}=\int \mathrm{d}\mb{K} C(\mb{K}) \exp{(-\mathrm{i}\omega_\mb{K}t)}\ket{\mb{K}}=\int \mathrm{d}\mb{K} C(\mb{K}) \exp{(\mathrm{i}\mb{K}\cdot\mb{R}-\mathrm{i}\omega_\mb{K}t)}
\end{equation}
where $\mathbf{K}=\mathbf{k}_1+\mathbf{k_2}$ is the centre-of-mass wavevector (see also Appendix~\ref{sec:complete-solution}) and $C(\mb{K})$ is the overlap integral of the eigenfunction and the initial state. To act with the field operators we expand the two-particle wave function $\Psi(\mb{r}'_1,\mb{r}'_2,t)$ into the Fock basis, 
\begin{equation}
    \ket{\Psi(t)}=\sum_{f,f'}
    \frac{1}{\sqrt{2}} \int \mathrm{d}^3r'_1 \mathrm{d}^3r'_2 \Psi(\mb{r}'_1,\mb{r}'_2,t) \hat{\Psi}^\dagger_{f}(\mb{r}'_1) \hat{\Psi}^\dagger_{f'}(\mb{r}'_2) \ket{0}
    \label{eq:TwoParticleStateFock2}
\end{equation}
\subsection{Action of the field operators}
Acting with two annihilation operators on the state Eq.~\eqref{eq:TwoParticleStateFock2} evaluates the wavefunction at the detector coordinates, which can be calculated to (explicitly shown in Subsection~\ref{sec:ActingOnState2} below)
\begin{equation}
    \hat{\Psi}_{s'}({\mb{r}_2})\hat{\Psi}_{s}(\mb{r}_1)\ket{\Psi(t)}=\frac{1}{\sqrt{2}}\left(\Psi_{s,s'}(\mb{r}_1,\mb{r}_2,t)-\Psi_{s',s}(\mb{r}_2,\mb{r}_1,t)\right)\ket{0}
    \label{eq:Action_ON_TPS2}
\end{equation}
where the relative minus sign follows directly from the anti-commutation relations of the field. It is here where the exchange symmetry, discussed in the main text and imposed by hand in Eq.~\eqref{ANTISYM}, arises from quantum mechanical algebra. Using $\mb{r}_1=\mb{R}+\mb{r}/2$ and  $\mb{r}_2=\mb{R}-\mb{r}/2$ and Eq.~\eqref{Separation2}, the bracket corresponds to the antisymmetrized wavefunction 
of Eqs.~\eqref{eq:PSI_AS_Product} and~\eqref{eq:phi_AS}, albeit with the spins still in the product base. The change to the coupled spin basis is done in the following subsection.
\subsection{Reduction of the spin sum}
When calculating Eq.~\eqref{eq:G2_SUM2} the incoherent sum over the spins $s,s'$ is in fact a trace over the two-particle spin space in the product basis $\{s,s'\}$ and therefore invariant under the unitary transformation to the coupled basis $\{S,M\}$, i.e.
\begin{equation}
    \sum_{s,s'}|\cdot|^2
    =\mathrm{Tr}_{\mathrm{spin}}(\cdot)
    =\sum_{S,M}|\cdot|^2
    =\sum_{S}g_S|\cdot|^2
    =3|\cdot|_{S=1}^2+1 |\cdot|_{S=0}^2
\end{equation}
where $S=0,1$ denotes the singlet/triplet state and $g_S=2S+1$ is the multiplicity of the spin. The last equality follows because the triplet sector $S=1$ contains
three spin projections, $M_S=-1,0,1$, whereas the singlet sector $S=0$
contains only one spin projection $M_S=0$.
Plugging this result into Eq.~(30), we obtain
\begin{equation}\label{eq:app-G-2-spins}
    G^{(2)}(\mb{r}_1,\mb{r}_2)=\sum_{s,s'} |\psi(\mathbf{R},t) \Phi_{s,s'}(\mathbf{r},t)|^2=\sum_S g_S|\psi(\mathbf{R},t) \Phi_{S}(\mathbf{r},t)|^2=\sum_Sg_S|\Psi_{S}(\mathbf{R},\mb{r},t)|^2
\end{equation}
where the spatial function is symmetrized/antisymmetrized accordingly. Eq.~\eqref{eq:app-G-2-spins} corresponds to Eq.~\eqref{eq:G2LinkedToWF} in the main text, with the origin of the spin degeneracy made more explicit. 

\subsection{Action of the field operators on a two-particle state}\label{sec:ActingOnState2}
For completeness, let us explicitly derive Eq.~\eqref{eq:Action_ON_TPS2}. Let $\ket{\Phi}$ be the fermionic two-particle state of Eq.~\eqref{eq:TwoParticleStateFock2}. Acting with our field operators on that state leads to:
$$\begin{aligned}
    &\hat{\Psi}_{s'}(\mb{r}_2) \hat{\Psi}_s(\mb{r}_1) \ket{\Phi}=\sum_{f,f'} \hat{\Psi}_{s'}(\mb{r}_2) \hat{\Psi}_s(\mb{r}_1) \frac{1}{\sqrt{2}}\int \mathrm{d}^3 r_A \mathrm{d}^3 r_B \Phi(\mb{r}_A,\mb{r}_B) \hat{\Psi}^\dagger_{f}(\mb{r}_A) \hat{\Psi}^\dagger_{f'}(\mb{r}_B) \ket{0}\\
    &=\sum_{f,f'}\hat{\Psi}_{s'}(\mb{r}_2) \frac{1}{\sqrt{2}} \int \mathrm{d}^3 r_A \mathrm{d}^3 r_B \Phi(\mb{r}_A,\mb{r}_B)(\delta_{s,f}\delta(\mb{r}_A-\mb{r}_1)-\hat{\Psi}^\dagger_{f}(\mb{r}_A)\hat{\Psi}_{s}(\mb{r}_1))\hat{\Psi}^\dagger_{f'}(\mb{r}_B) \ket{0}\\
    &=\sum_{f,f'}\hat{\Psi}_{s'}(\mb{r}_2) \frac{1}{\sqrt{2}} \int \mathrm{d}^3 r_A \mathrm{d}^3 r_B \Phi(\mb{r}_A,\mb{r}_B)[\delta_{s,f}\delta(\mb{r}_A-\mb{r}_1)\hat{\Psi}^\dagger_{f'}(\mb{r}_B)-\delta_{s,f'}\delta(\mb{r}_B-\mb{r}_1)\hat{\Psi}^\dagger_{f}(\mb{r}_A)]\ket{0}\\
    &=\sum_{f,f'}\hat{\Psi}_{s'}(\mb{r}_2) \frac{1}{\sqrt{2}} \left[\int \mathrm{d}^3 r_B \Phi(\mb{r}_1,\mb{r}_B)\delta_{s,f}\hat{\Psi}^\dagger_{f'}(\mb{r}_B)- \int \mathrm{d}^3 r_A \Phi(\mb{r}_A,\mb{r}_1)\delta_{s,f'}\hat{\Psi}^\dagger_{f}(\mb{r}_A)\right]\ket{0}\\
    &=\sum_{f,f'}\frac{1}{\sqrt{2}} \left[\int \mathrm{d}^3 r_B \Phi(\mb{r}_1,\mb{r}_B)\delta_{s,f}\delta_{s',f'}\delta(\mb{r}_B-\mb{r}_2)-\int \mathrm{d}^3 r_A \Phi(\mb{r}_A,\mb{r}_1)\delta_{s,f'}\delta_{s',f}\delta(\mb{r}_A-\mb{r}_2)\right]\ket{0}\\
    &=\sum_{f,f'}\frac{1}{\sqrt{2}} \left[\Phi(\mb{r}_1,\mb{r}_2)\delta_{s,f}\delta_{s',f'}- \Phi(\mb{r}_2,\mb{r}_1)\delta_{s,f'}\delta_{s',f} \right]\ket{0}\\
    &=\frac{1}{\sqrt{2}} \left[\Phi_{s,s'}(\mb{r}_1,\mb{r}_2)- \Phi_{s',s}(\mb{r}_2,\mb{r}_1)\right]\ket{0}
\end{aligned}$$
where we used the anticommutator relations and $\hat{\Psi}_i(\mb{r}_j)\ket{0}=0$.

\subsection{Remark: directional momentum states}
Note that in Eq.~\eqref{eq:InsertCompleteness} any complete set of eigenfunctions can be used. A convenient alternative, useful for numerical evaluations, is provided by the directional scattering states, labeled by the vector $\mb{k}$ and normalized such that they reduce to plane waves for vanishing Coulomb interaction $\eta\rightarrow0$. In terms of the radial function Eq.~\eqref{RADIALexact} they read~\cite{QuantumMechanics}
\begin{equation}
    \Phi^{(+)}_\mathbf{k}(\mathbf{r})=\frac{1}{k} \sum^\infty_{l=0}(2l+1) \mathrm{e}^{\mathrm{i}\delta_l} \mathrm{i}^l R_{k,l}(r) P_l(\mathbf{\hat{k}}\cdot\mathbf{\hat{r}})
    \label{DirectMomentumSpherical}
\end{equation}
This can also be written in closed form \cite{QuantumMechanics}
\begin{equation}
    \Phi^{(\pm)}_k(\mathbf{r})=\Gamma(1\pm\mathrm{i}\eta(k))\mathrm{e}^{\frac{-\pi \eta(k)}{2}} \mathrm{e}^{\mathrm{i}\mathbf{k}\cdot\mathbf{r}} \;_1F_1 (\mp\mathrm{i} \eta(k)|1|\pm \mathrm{i} k r-\mathrm{i}\mathbf{k}\cdot \mathbf{r}),
    \label{CarEigenfucntion}
\end{equation}
what avoids the partial sum appearing in Eq.~\eqref{psi_Workhorse} and thus makes the expression more efficient for numerical calculations. 

\section{Complete solution including center-of-mass system}\label{sec:complete-solution}

For completeness, we also present the solution of the center-of-mass system and the resulting complete solution.

\subsection*{Center-of-mass system}
The Schrödinger equation for the center-of-mass system reads
\begin{equation}
    \mathrm{i} \hbar  \frac{\partial}{\partial t} \psi (\mathbf{R},t)=-\hbar^2 \frac{\Delta_\mathbf{R}}{2 M} \psi (\mathbf{R},t) = \hat{H}_{\mathbf{R}} \psi (\mathbf{R},t) \, ,
\end{equation}
which is identical to the Schrödinger equation of a free particle with mass $M=2 m_e$, where $\Delta_\mathbf{R}$ denotes the Laplacian operator with respect to the center-of-mass coordinate $\mathbf{R}$.
The eigenfunctions are given by plane waves 
\begin{equation}
    \psi_\mathbf{K}(\mathbf{R}) = \exp(\mathrm{i} \mathbf{K}\cdot \mathbf{R})
\end{equation}
with time evolution
\begin{equation}
    \psi_\mathbf{K}(\mathbf{R},t)= \psi_\mathbf{K}(\mathbf{R}) \exp(-i\omega_K t)
\end{equation}
where $\omega_K = E_K/\hbar$ and $E_K=\hbar^2\mathbf{K}^2 /2 M$ with $\mathbf{K}=\mathbf{k}_1+\mathbf{k_2}$ being the centre-of-mass wavevector.

\subsection*{Complete solution}

The complete solution of the two-electron Schrödinger equation of Eq.~\eqref{SchrödingerEQ_Normal} is then a linear combination of the combined eigenfunctions
\begin{equation}
    \Psi_{\mathbf{K},k,l,m}(\mathbf{R},\mathbf{r},t)=\psi_\mathbf{K}(\mathbf{R},t) \Phi_{k,l,m}(\mathbf{r},t) \, ,
\end{equation}
combining the solutions of the previous section with the solution of the relative system in the main text. 

Knowing the initial state of the two-electron system, $\Psi(\mathbf{R},\mathbf{r},t=0)$, we can write the general solution in terms of these eigenfunctions of the scattering Hamiltonian via
\begin{equation}
    \Psi(\mathbf{R},\mathbf{r},t)_{}=\int \mathrm{d}\mathbf{K} \int \mathrm{d}{k} k^2 \sum^{\infty}_{l=0} \sum^l_{m=-l} \zeta_{\mathbf{K},k,l,m} \Psi_{\mathbf{K},k,l,m}(\mathbf{R},r) \mathrm{e}^{-\mathrm{i}\omega_k t} \mathrm{e}^{-\mathrm{i}\omega_K t},
    \label{Formal Solution}   
\end{equation}
where 
\begin{equation}
    \zeta_{\mathbf{K},k,l,m}=\int_{\Omega_r} \int_{\Omega_R} \Psi (\mathbf{R}, \mathbf{r},t=0) 
    \Psi^*_{\mathbf{K},k}(\mathbf{R},\mathbf{r})
    \mathrm{d}\mathbf{r} \mathrm{d} \mathbf{R},
    \label{OverlapIntegral}
\end{equation}
is the overlap integral of the initial state and the spatial eigenfunctions of the scattering Hamiltonian. 

\section*{References}

\bibliographystyle{iopart-num}
\bibliography{ref}

\end{document}